\documentclass[aip,rsi,reprint,twocolumn]{revtex4-1}
\usepackage[pdftex]{graphicx}
\usepackage{hyperref}
\usepackage{url}
\usepackage{dcolumn}
\usepackage{bm}
\usepackage{amsmath}
\usepackage{amssymb}

\newcommand{\mrm}{\mathrm}

\begin{document}
\title{Laser frequency stabilization using a transfer interferometer}
\author{Shira Jackson}
\author{Hiromitsu Sawaoka}
\author{Nishant Bhatt}
\author{Shreyas Potnis}
\author{Amar C. Vutha}
\affiliation{Department of Physics, University of Toronto, Toronto ON M5S 1A7, Canada}
\email{amar.vutha@utoronto.ca}

\begin{abstract}
We present a laser frequency stabilization system that uses a transfer interferometer to stabilize slave lasers to a reference laser. Our implementation uses off-the-shelf optical components along with microcontroller-based digital feedback, and offers a simple, flexible and robust way to stabilize multiple laser frequencies to better than 1 MHz.
\end{abstract}

\maketitle

\section{Introduction}
Laser frequency stabilization is an essential element of almost every experiment in atomic and molecular physics. The usual approach to stabilizing the frequency of a laser is to lock it to an atomic or molecular resonance (e.g.,\cite{Ma2007,Gong2014, Kobayashi2015}). However, this approach is limited to lasers with frequencies that are within a small neighborhood of a conveniently accessible atomic or molecular transition. A common alternative is to stabilize a laser to the resonances of a Fabry-Perot cavity, using the Pound-Drever-Hall method \cite{Drever1983} or frequency-modulation techniques \cite{Bjorklund1983}. A single cavity can also be used to stabilize multiple lasers, either using acousto-optic or electro-optic modulators to tune the lasers into resonance with a fixed cavity, or by using a transfer cavity and a stable reference laser (RL) to stabilize a number of slave lasers (SLs) \cite{Zhao1998,Rossi2002,Burke2005,Matsubara2005,Bohlouli-Zanjani2006,Rohde2010,Seymour2010,Tonyushkin2010,Yang2012}. In a common variation of the basic scheme, the stability transfer is accomplished by continuously scanning the length of an optical cavity over at least one free spectral range (FSR), and stabilizing the positions of the SL transmission peaks relative to the position of the RL transmission peak, using analog or digital feedback. Laser frequency stabilization to a level $\lesssim$1 MHz has been reported using variations of this technique. 

An advantage of the transfer cavity method is that multiple lasers can be stabilized at arbitrary frequencies that can be far from atomic or molecular resonances. Another advantage is that it results in a wide capture range, on the order of the cavity FSR, which allows the stabilization system to recover from large perturbations of the SL frequencies. However, when the RL and SLs have very different wavelengths, customized optical cavity mirrors with multi-wavelength high-finesse coatings are usually necessary for constructing a transfer cavity. This also means that it is not always easy to add new lasers to an existing cavity-based stabilization system.

Here we present a simple alternative method for laser frequency stabilization based on a transfer \emph{interferometer} instead of a cavity, which can be rapidly implemented using off-the-shelf optical components. The interferometer-based method lends itself to a modular implementation that is significantly easier to assemble compared to a transfer cavity, but nevertheless exhibits a level of performance that is comparable to more sophisticated transfer cavity schemes. It also inherits the advantages of the transfer cavity method, making it a convenient choice for stabilizing a family of cooling and repump lasers for atomic or molecular laser-cooling, for example. We describe the theory of operation of our laser stabilization method in Section \ref{sec:theory}, and present details of the implementation and measured performance in Section \ref{sec:implementation}.

\section{Theory of operation}\label{sec:theory}
\begin{figure}
\centering
\includegraphics[width=0.7\columnwidth]{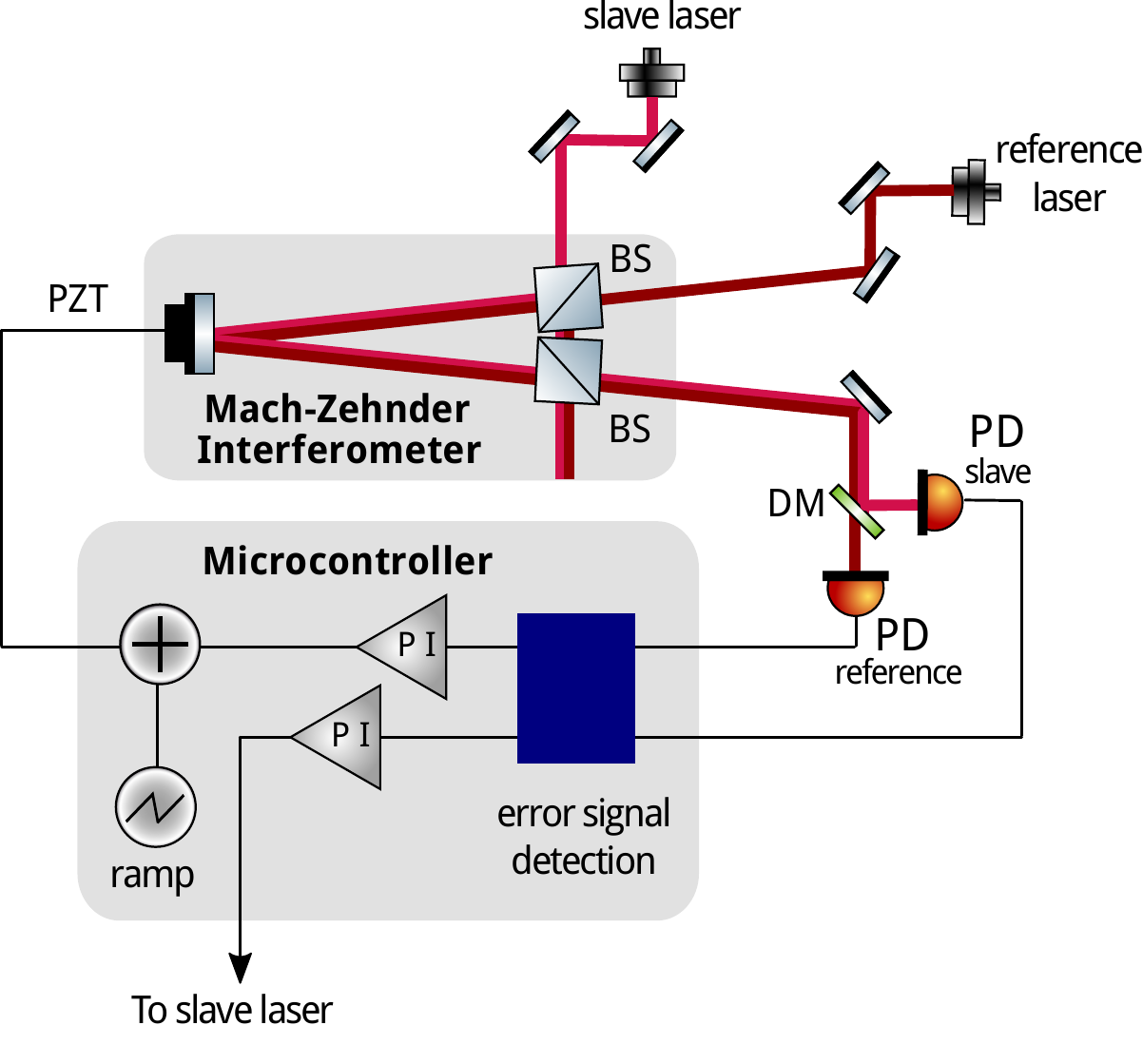}
\caption{Schematic of the setup for laser frequency stabilization. The schematic is shown with a single slave laser for simplicity, but the method can be used to stabilize multiple slave lasers with the same transfer interferometer. The abbreviations used are: PZT, piezoelectric actuator; BS, beamsplitter; DM, dichroic mirror; PD, photodiode; PI, proportional-integral feedback loop filter.}
\label{fig:schematic}
\end{figure}

In our method, the stability of a reference laser (RL) is transfered to a number of slave lasers (SLs) using a scanning interferometer. A schematic of the transfer interferometer-based laser stabilization (TILS) system is shown in Fig.\ \ref{fig:schematic}. Information about laser frequency drifts is derived from the phase of the sinusoidal fringe signals (as shown in Fig.\ref{fig:signals}), which are observed as the path length difference, $L$, of the interferometer is scanned. The phase of a slave laser fringe signal, $\phi_S$, is determined from a least squares fit to the fringe signal measured on a photodiode. The phase $\phi_S$ is combined with the phase of the reference laser fringe signal, $\phi_R$, to obtain an error signal that is independent of fluctuations in the length of the interferometer, $\delta L$, to leading order. With $k_R \, (k_S)$ denoting the wavevector of the reference (slave) laser, the equations for first-order fluctuations of the phases are
\begin{equation}\begin{split}
\delta \phi_R & = k_R \, \delta L \\
\delta \phi_S & = \delta k_S \, L + k_S \, \delta L.
\end{split}\end{equation}
Therefore the error signal $e_S \propto \left( \frac{\delta \phi_S}{k_S} - \frac{\delta \phi_R}{k_R}\right)$ estimates the frequency deviation of the SL in a way that is independent from $\delta L$. This error signal is used in a feedback loop to stabilize the frequency of the slave laser. Such a phase-based error signal is also conveniently insensitive to laser power fluctuations, and to drifts in the offsets of the detectors or the interference fringe contrast. Since the phase can be unambiguously determined over the range $(-\pi,\pi)$ corresponding to one FSR of the interferometer, the TILS method yields a large lock capture range similar to other transfer-cavity-based schemes, allowing it to robustly recover from large perturbations of the laser frequencies. 

The stability of the locked slave laser's frequency, $\Delta \nu_S$, can be estimated from the phase uncertainty of the least squares fit. For a sinusoid with signal to noise ratio $\mathcal{S}$ that is sampled at $N$ points, the phase uncertainty from a least squares fit is $\Delta \phi = \sqrt{\frac{2}{N}} \frac{1}{\mathcal{S}}$ \cite{Montgomery1999}. Therefore the frequency stability of a locked slave laser (assumed to be limited only by this measurement error) is
\begin{equation}\label{eq:delta_nu}
\Delta \nu_S = \sqrt{\frac{2}{N}} \frac{c}{2 \pi n(\nu_S) L \mathcal{S}},
\end{equation}
where $n(\nu_S)$ is the refractive index at the frequency of the slave laser. It is instructive to compare Equation (\ref{eq:delta_nu}) with the frequency uncertainty of a laser locked to a cavity with finesse $\mathcal{F}$, round-trip length $L$ and the same detection SNR $\mathcal{S}$, which is equal to $(\Delta \nu_S)_\mrm{cav} =  \frac{c}{\mathcal{F} n(\nu_S) L \mathcal{S}}$. It is evident that locking to an interferometer is equivalent to using a cavity with finesse $\mathcal{F}_\mrm{eff} = \pi \sqrt{2N}$. The intuitive reason for this extra enhancement factor, compared to the naively expected $\mathcal{F} \sim 1$, is that half the number of samples on a sinusoidal fringe (on average) are useful for determining the frequency shift of the laser, whereas a scanning cavity only yields frequency shift information in the vicinity of the cavity resonance. 

\section{Implementation}\label{sec:implementation}
In our implementation, the RL is a compact interference-filter-based external-cavity diode laser (ECDL) whose frequency is stabilized to an atomic transition in rubidium via sub-Doppler dichroic atomic vapor laser lock (SD-DAVLL) \cite{Potnis2017}. For the measurements characterizing the interferometer locking scheme that described below, the SL is an 852 nm ECDL. 

The transfer interferometer (see Fig.\ref{fig:schematic}) has one short arm and one long arm, with a path length difference of $L=26$ cm. If needed, it is straightforward to increase the path difference to obtain a more sensitive discriminator signal (e.g., using an optical fiber). The interferometer is constructed using off-the-shelf mirrors and beamsplitters, and is mounted inside an aluminum enclosure with a temperature-stabilized baseplate. A mirror mounted on a piezoelectric stack is used to scan the length of the interferometer. The RL and SL beams are coupled into the same interferometer. In order to suppress long-term drifts due to misalignment of the laser beams, it is best to propagate the RL and SL along the same path through the interferometer. This can be implemented using an optical fiber or pinhole pairs to ensure collinear propagation of the RL and SL beams.

\begin{figure}
\centering
\includegraphics[width=\columnwidth]{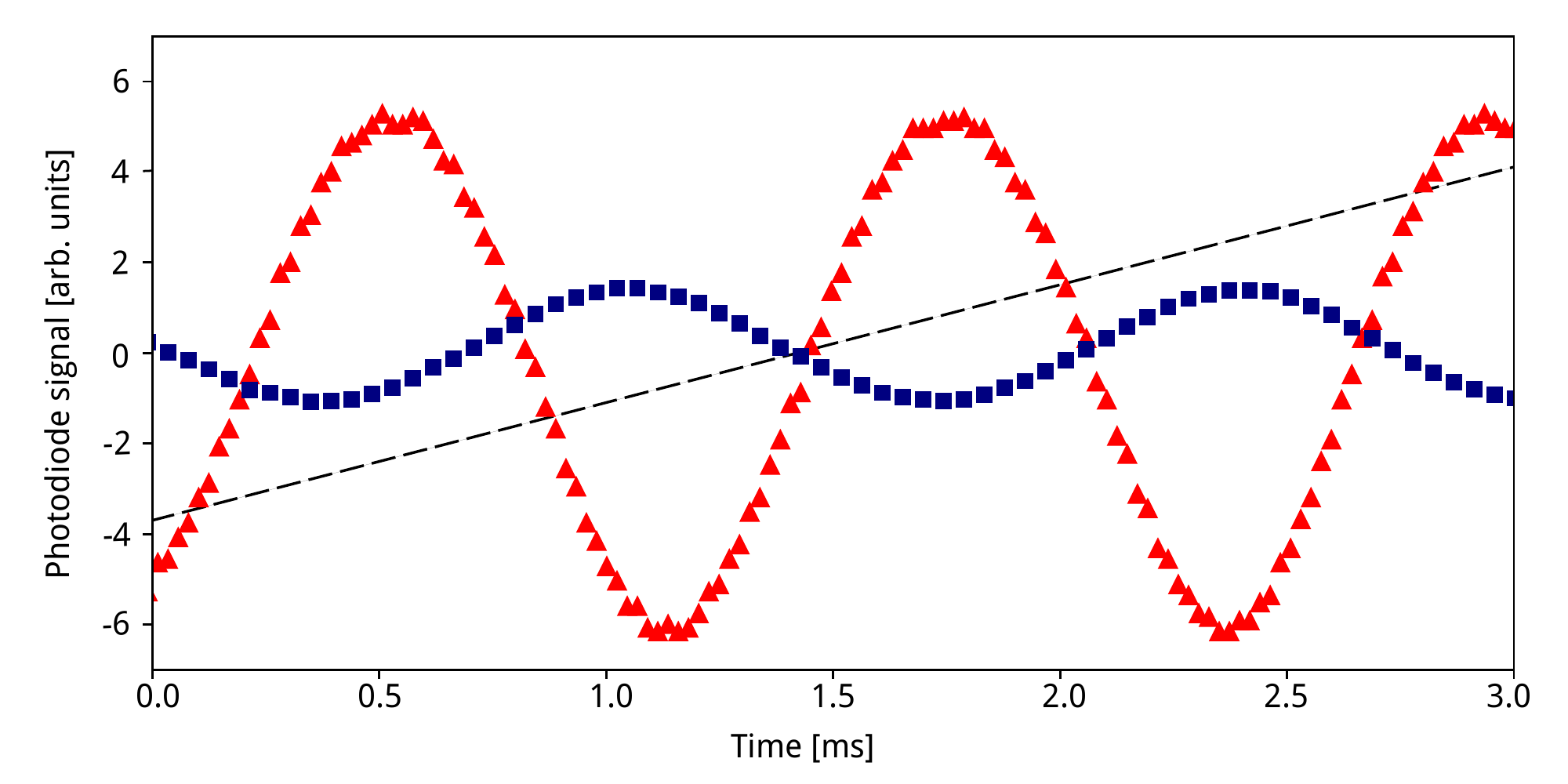}
\caption{The voltage ramp applied to scan the interferometer (indicated as a dashed line), along with typical interference fringe signals from the reference laser (RL, red triangles) and slave laser (SL, blue squares). The phases of the fringe signals from the SL and RL are processed by digital feedback loops to stabilize the SL frequency and the average length of the interferometer.}
\label{fig:signals}
\end{figure}

The output of the interferometer is separated using dichroic mirrors, and each laser is sent into a separate photodiode. The photodiode signals are processed using a microcontroller, which is also used to generate the voltage ramp for scanning the interferometer length, as well as to generate correction signals for stabilizing the SLs and the average length of the interferometer. High signal-to-noise ratio fringes are readily obtained from the interferometer without the need for amplified photodiodes, as seen in Fig.\ \ref{fig:signals}.

Our implementation uses an Arduino-compatible microcontroller\footnote{Commercial products are identified for the sake of clarity -- this does not constitute an endorsement.}, along with an analog-digital converter (ADC) board\footnote{Digilent Analog Shield, \url{http://reference.digilentinc.com/analog_shield}} for reading in the photodiode signals and applying corrections to the interferometer and SLs. The interferometer is typically scanned with a sawtooth waveform at $f_\mrm{scan} = 200$ Hz over a couple of fringes. The fringes measured on the photodiodes are digitized at a sampling rate of $\sim$ 5 kS/s. The phase of each sinusoidal fringe is computed using a linear least squares routine \cite{Handel2000} implemented in the microcontroller. As described in Section \ref{sec:theory}, least squares detection of the fringe phase shifts efficiently uses multiple sample points over the fringe, and is practically more robust as well compared to detection of just the zero crossings. The error signal $e_S$ is fed into a proportional-integral (PI) controller and the resulting control signal is used to adjust the frequency of the SL. The lock point for $\phi_S$ can be easily changed in software over the range $(-\pi,\pi)$ to scan the SL over one FSR of the interferometer, $c/L \approx$ 1.2 GHz.

While the error signal $e_S$ described in Section \ref{sec:theory} is insensitive to small fluctuations in $L$, thermal and other drifts in $L$ that shift the interferometer's output by one or more fringes will lead to drifts in the lock points. Therefore we also stabilize the average length using an error signal proportional to $\phi_R$, which is sent to a PI-controller that adjusts the bias voltage on the interferometer piezo. 

The microcontroller code and a parts list are available online \footnote{\url{http://github.com/vuthalab/Transfer-Interferometer}}. The loop bandwidth of the TILS system is determined by the sum of the scan period, analog-digital and digital-analog conversion times and processing time. If required, it is straightforward to improve the processing time using separate microcontrollers for each laser or a field-programmable gate array (e.g.,\cite{Yang2012}), and to improve the scan rate using a faster piezoelectric actuator \cite{Briles2010} and a faster ADC. 
\begin{figure}[tp]
\centering
\includegraphics[width=\columnwidth]{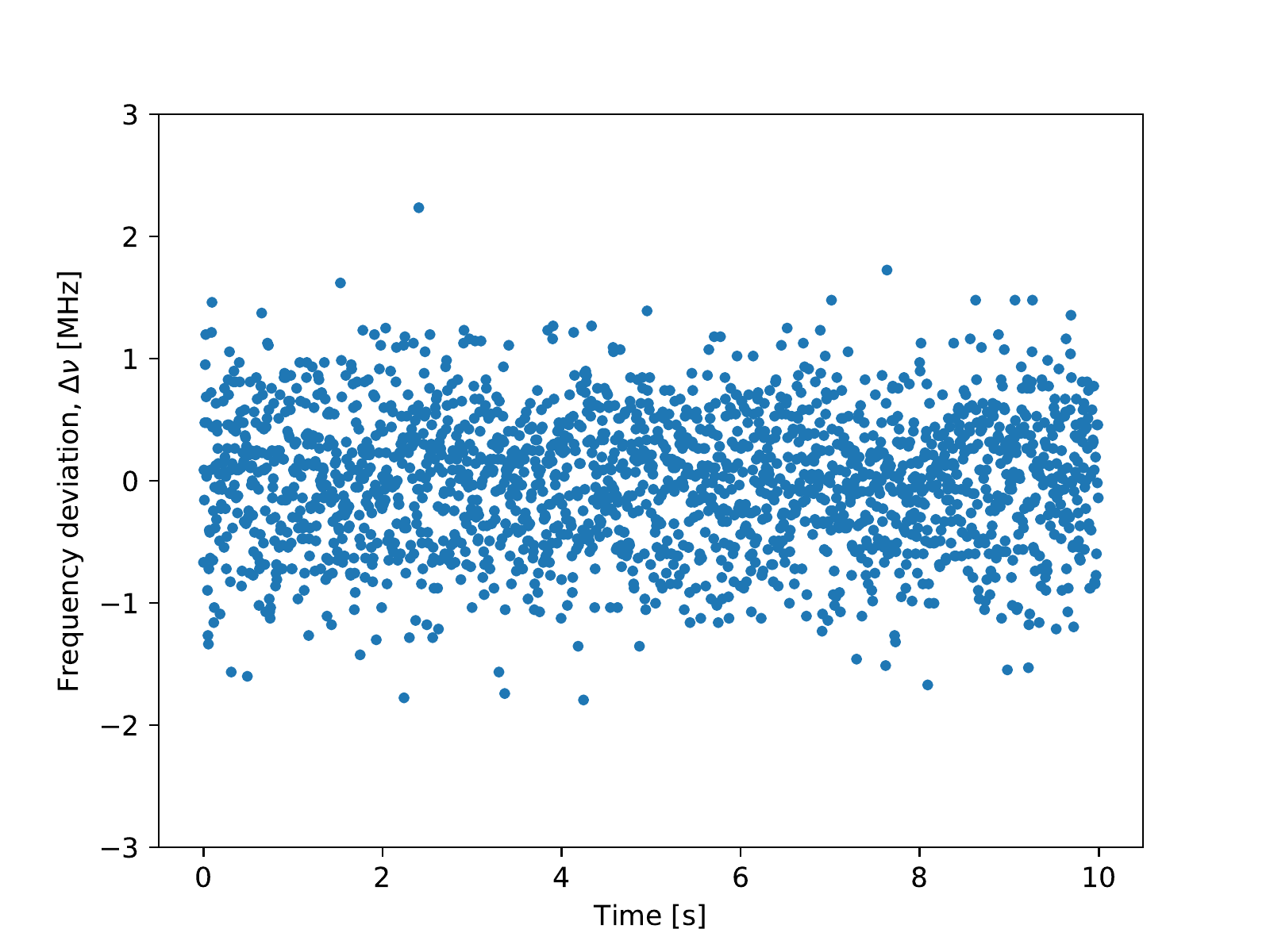} 
\caption{Short-term stability of the TILS system, measured using a stabilized 852 nm laser. The rms frequency deviation over the 10 s interval shown above is 550 kHz.} \label{fig:short_term_stability}
\end{figure}
\begin{figure}[h!]
\centering
\includegraphics[width=\columnwidth]{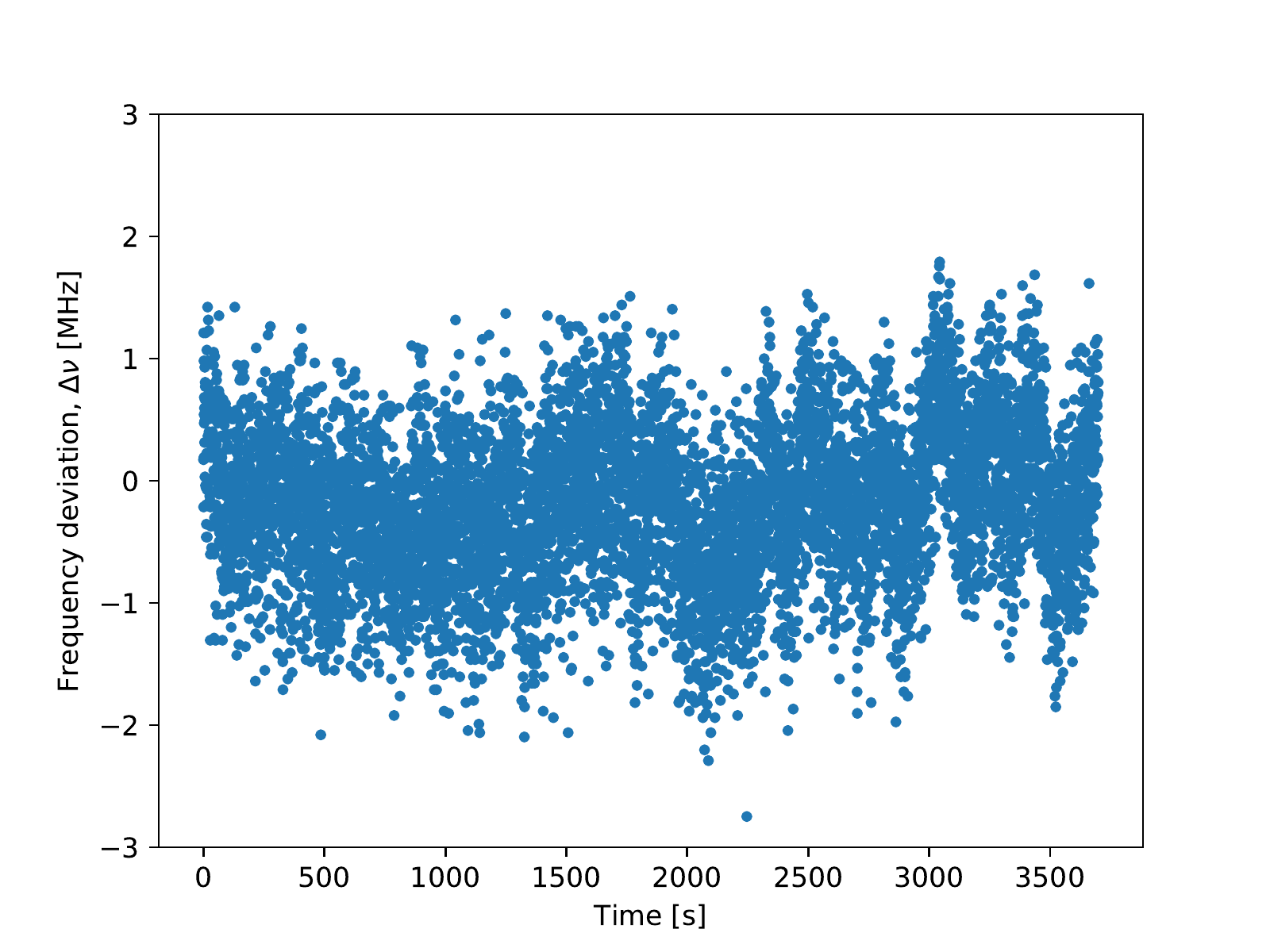}
\caption{Long-term stability of the TILS system, measured using a stabilized 852 nm laser. Over the $\sim$1 hr span shown above, the rms frequency deviation is 620 kHz.} \label{fig:long_term_stability}
\end{figure}

The predicted SL stability from Equation (\ref{eq:delta_nu}), using typical numbers for our apparatus ($N=8$, $L=26$ cm and $\mathcal{S} \approx 150$), is $\Delta \nu_S \approx$ 0.6 MHz. In order to evaluate the performance of the TILS system, the 852 nm laser was separately frequency-stabilized to a cesium vapor cell. The measured phase fluctuations were converted to laser frequency deviations and logged for analysis. Since the 780 nm and 852 nm lasers fluctuate while locked to their respective atomic resonances (typically by $\sim$100 kHz), the frequency deviations reported here represent an upper bound on the fluctuations in the TILS system. Figs.\ \ref{fig:short_term_stability} and \ref{fig:long_term_stability} show the frequency deviations measured on different timescales. On both the short ($\sim$10 s) and longer ($\sim$1 hr) timescales, the rms frequency deviation remains below 1 MHz, and is at a level that is consistent with the predicted stability. 

We have used this interferometer-based system to stabilize multiple lasers at different wavelengths (e.g., 423 nm and 453 nm lasers for cooling Ca atoms, stabilized to an 780 nm reference laser) and observed stable performance over many days without the need for realignments.

In summary, we have described the construction and performance of a simple and modular laser stabilization system that combines ease of implementation with good stability. We acknowledge helpful feedback from Andrew Jayich. This work is supported by funding from NSERC, CFI and Society in Science.

\bibliography{laser_stabilization} 

\end{document}